\documentclass[aps,showpacs,showkeys,preprint]{revtex4}
\usepackage{epsbox}
\usepackage[dvips]{color}
\begin{document}
%\textsf{epsbox}
% Use the \preprint command to place your local institutional report
% number in the upper righthand corner of the title page in preprint mode.
% Multiple \preprint commands are allowed.
% Use the 'preprintnumbers' class option to override journal defaults
% to display numbers if necessary
\preprint{KOBE-TH-03-07}

%Title of paper
\title{Protecting the primordial baryon asymmetry in the seesaw model compatible with WMAP and KamLAND}

% repeat the \author .. \affiliation  etc. as needed
% \email, \thanks, \homepage, \altaffiliation all apply to the current
% author. Explanatory text should go in the []'s, actual e-mail
% address or url should go in the {}'s for \email and \homepage.
% Please use the appropriate macro foreach each type of information

% \affiliation command applies to all authors since the last
% \affiliation command. The \affiliation command should follow the
% other information
% \affiliation can be followed by \email, \homepage, \thanks as well.

\author{~K.~Hasegawa}
\email[]{kouhei@phys.sci.kobe-u.ac.jp}
\affiliation{Department of Physics, Kobe University, Rokkodaicho 1-1,
Nada ward, Kobe 657-8501, Japan}
%\homepage[]{Your web page}
%\thanks{}
%\altaffiliation{}

%Collaboration name if desired (requires use of superscriptaddress
%option in \documentclass). \noaffiliation is required (may also be
%used with the \author command).
%\collaboration can be followed by \email, \homepage, \thanks as well.
%\collaboration{}
%\noaffiliation

\date{\today}

\begin{abstract}
We require that the primordial baryon asymmetry is not washed out in the seesaw model 
compatible with the recent results of WMAP and the neutrino oscillation experiments including 
the first results of KamLAND. We find that only the case of the normal neutrino mass hierarchy with 
an approximate $L_{e}$-symmetry satisfies the requirement. We further derive, depending
on the signs of neutrino mass eigenvalues, three types of neutrino mass matrixes, where
the values of each element are rather precisely fixed.
\end{abstract}
 
% insert suggested PACS numbers in braces on next line

% insert suggested keywords - APS authors don't need to do this

%\maketitle must follow title, authors, abstract, \pacs, and \keywords
\maketitle

% body of paper here - Use proper section commands
% References should be done using the \cite, \ref, and \label commands
\section{Introduction\label{intro}}
The results of the recent neutrino oscillation experiments showed that neutrinos have non-zero masses 
\cite{S-K, S-K2, Valle-Gon, SNO, SNO2, kam, kam2, Chooz}. 
The global analysis of the very recent first results of KamLAND combined with the other existing neutrino experiments
 gives us the more precise information about the neutrino mass matrix \cite{Valle-kam}.
 Although we don't know the magnitudes themselves of the neutrino mass eigenvalues  from 
 the neutrino oscillation experiments, the investigation into the anisotropy of the cosmic microwave
 background radiation (CMBR) by Wilkinson Microwave Anisotropy Probe (WMAP) recently has put
 the upper bounds on the sum of the neutrino masses \cite{Wmap1,Wmap2,Han,Elg,Wmap3}.
 There are works that the neutrino mass matrixes are analyzed by these recent results of KamLAND
 and  WMAP \cite{Bha,Kin,Zee,Joa}.
 Thus the search for physics beyond the Standard Model, which incorporates non-zero neutrino masses,
 has become an urgent issue.
 Seesaw model is one of the most promising candidates for such neutrino mass generation \cite{Yan, Yan2, Yan3}.
 Seesaw model also works as the model of leptogenesis, which can explain the present baryon number
  through the sphaleron processes \cite{Fuk, Pil, Ham}.
 There are many works about the relation between the generated baryon number
 at higher temperatures and the neutrino mass matrix or CP phases 
  at the low temperature universe \cite {Buc1, Buc2, Chu, Nie, Jos, Hir, Bra, End,  Bal, Bra2, Dav, Fal, Kan, Kan2}.
    On the other hand, in the seesaw model the lepton number is inevitably broken by the Majorana mass term,
	and the model potentially erases the baryon and lepton numbers when combined with the sphaleron processes
 in equilibrium \cite{Fuk2, Hav, Nel, Buc3,  Fis, Oli1, Oli2, Dre}. It has been argued that 
 the condition to protect the baryon number from this washing out processes gives the stringent 
   constraints for the dimension five operator including Higgs fields,
   which in turn induces the Majorana neutrino masses at symmetry broken phase.
  In the present paper, inspired by the very recent experimental progress in neutrino physics, i.e.
the KamLAND and WMAP, we re-analyze the condition for the baryon and lepton numbers
not to be washed out by the decay and the inverse-decay processes of heavy right-handed
neutrinos and by the dimension five operator. We require the condition, irrespectively of what kind
of theory provides the primordial baryon or lepton asymmetry at higher temperatures, and 
will see whether there remains some allowed parameter range of neutrino mass matrix.
When the magnitudes of the six independent CP violating phases \cite{End}
are not enough for the leptogenesis through the out-of-equilibrium
decays of the right-handed neutrinos \cite{Fuk} to provide the present baryon number, 
the investigation extended here becomes important. From this viewpoint, we assume that both 
the Yukawa couplings and Majorana mass matrix of the right-handed neutrinos are all real.
Then, the real Majorana mass matrix of left-handed neutrinos has the six independent
 parameters. On the other hand, the neutrino oscillation experiments determine the five parameters, two
squared mass differences and three mixing angles.
We thus can describe the neutrino mass matrix in terms of only one real parameter, $m_{1}$ or $m_{3}$
within the error of the experimental data.
Under this assumption, we require in the seesaw model the conditions that
 the decay and inverse-decay processes of the heavy Majorana neutrino and the processes induced from the
 dimension five operator are out-of-equilibrium in order to protect the  baryon asymmetry
 and study whether such constrained seesaw model can simultaneously satisfy the results of the neutrino oscillation
  experiments including KamLAND and those of WMAP or not.
 We can further determine the neutrino mass matrix with accuracy.
  The following criticism of this assumption may be thought of: 
  the mass scale of the right-handed Majorana neutrinos is expected to be very high, for example, GUT scale.
  Even if the decay and inverse-decay processes of the heavy Majorana neutrino are in equilibrium and all the lepton and
   baryon numbers are washed out in the higher scale, the possibility of the various baryogenesis may be left at the lower scale.
   But we can exclude such possibility by making the mass scale of 
   right-handed Majorana neutrinos relatively low, for example, TeV scale.
   We can realize this when the Dirac Yukawa couplings of  right-handed neutrinos are set to be small enough
 or the descent of the rank of the neutrino mass matrix explains the smallness of left-handed neutrino masses as
 was recently pointed out in \cite{Loi}.
 
 If we require that all the decay and inverse-decay processes of the heavy Majorana neutrinos are out-of-equilibrium
 to protect the baryon number,
 it turns out that the left-handed Majorana masses are all restricted  to the very small \cite{Buc3, Buc2},
 \begin{eqnarray}
 |m_{\nu}| \lesssim 10^{-3} [\mbox{eV}].
 \end{eqnarray}
 These small neutrino masses are clearly incompatible with the squared mass differences
 needed to account for the data of the neutrino oscillation.
 We may relax the condition and require that only one flavor symmetry
 among the three symmetries, $L_{e}, L_{\mu}$ and $L_{\tau}\ (L=L_{e}+L_{\mu}+L_{\tau})$,
 should be an exact symmetry. However, if the seesaw model has the $L_{e}$-symmetry, for example, it is easily shown 
 that the two of three mixing angles vanish, which contradicts with the results of the neutrino oscillation experiments.
Let us note the symmetry needs not to be an exact one and an approximate symmetry will
suffice for the purpose of protecting the baryon asymmetry.
  An approximate symmetry means that the rates of all the processes breaking the symmetry are smaller than
 the Hubble parameter. On the other hand, there
 are many studies which attempt to determine the Majorana neutrino mass matrix purely from 
 the neutrino oscillation data under the assumption of the two-zero textures \cite{Fra, Xin, Des, Bar, Hon}.
 They showed that the results of the neutrino oscillation experiments favor the mass matrixes
 with normal hierarchy of mass eigenvalues, which have the small or zero
elements $m_{e \alpha}(\alpha=e, \mu, \tau)$ and the inverted hierarchy is
 allowed for only one case, $m_{\mu \mu}=m_{\tau \tau}=0$. 
 These results imply that  the approximate  $L_{e}$-symmetry in the normal mass hierarchy is favored 
 and the inverted hierarchy is disfavored if one of the approximate symmetries is to be required 
 as a symmetry to protect the baryon asymmetry, being consistent with the data of all neutrino oscillation experiments.
  We will verify that it is really the case till the end of the present paper.
  
The outline of this paper is as follows: in Section II, we first discuss the condition to protect 
baryon asymmetry in the seesaw model. Next, we require the additional condition imposed by
the recent results of WMAP. In Section III, 
we study whether such constrained seesaw model can be compatible with the data of the neutrino oscillation experiments
 including the first results of KamLAND. In the argument we divide the cases into two, i.e. the normal and inverted
 mass spectra including quasi-degenerate type and rather precisely determine the neutrino mass matrix 
 depending on the signs of the mass eigenvalues.
 We next show that there exists a parameter set of the theory which corresponds to
 the allowed neutrino mass matrix. We also confirm that the allowed neutrino mass matrix satisfies the constraint from 
  the absence of the neutrinoless double beta decay ($0\nu\beta\beta$). We finally calculate the resultant
  lepton and baryon numbers in the present universe. Section IV is devoted  to a summary.

\section{Protecting the primordial baryon asymmetry in the seesaw model compatible with the results of WMAP }
In this section, we briefly review the seesaw model and discuss the condition that 
the primordial baryon asymmetry is not washed out. We further require that the seesaw model is 
compatible with the recent results of WMAP.   
\subsection{Brief review of the seesaw model}
In addition to the one in the Standard Model, 
the following Lagrangian density is included in the seesaw model,  
\begin{eqnarray}
{\cal L}^{FY}=\bar{N}_{R}^{i}i \partial \!\!\!/ N_{R}^{i}-\frac{1}{2}
\bigl( (\overline{N_{R}^{i})^{c}} \hat{M}_{R}^{i}N_{R}^{i}+\mbox{h.c.}\bigr)
+(h^{\alpha i}\bar{l}_{L}^{\alpha}\tilde{\Phi}N_{R}^{i} +\mbox{h.c.}),
 \end{eqnarray}
where the heavy right-handed neutrino $N_{R}^{i} (i=1, 2, 3)$ is the mass eigenstate with the mass eigenvalue 
$M_{R}^{i}$. The doublets of weak eigenstate $l_{L}^{\alpha}  (\alpha=e, \mu, \tau)$ 
are in the base where the mass matrix of the charged leptons
is diagonal. Namely, the lepton doublets and higgs doublet are defined as
\begin{eqnarray}
l_{L}^{\alpha}\equiv\left(\begin{array}{c}
                                     {\nu}_{\alpha} \\
                                      {e}_{\alpha}  
									 \end{array}\right)_{L}
									 \equiv\left(\begin{array}{ccc}
                                    {\nu}_{e}  &   {\nu}_{\mu}   &   {\nu}_{\tau} \\
                                     e  &\mu & \tau
									 \end{array}\right)_{L} \mbox{and} 
									 \ \tilde{\Phi} \equiv \left(\begin{array}{c}
                                      \phi^{0 \ast} \\
									  -\phi^{-} 
									 \end{array}\right).
									 \end{eqnarray}
Let us note that, without loss of generality, the right-handed Majorana mass matrix may
be assumed to be diagonal. The exchange of the heavy right-handed Majorana neutrinos provides
the dimension five operator,  breaking the lepton number by two units,
 \begin{eqnarray}
{\cal L}^{eff}=\frac{1}{2}\Bigl[H^{\ast} \frac{1}{\hat{M}_{R}} H^{\dagger}\Bigr]^{\alpha \beta}
(\overline{{\nu}_{L}^{\alpha})^{c}} \nu_{L}^{\beta}\phi^{0}\phi^{0}, \label{lfive}
 \end{eqnarray}
 where $(H)^{\alpha i}\equiv h^{\alpha i} \  \mbox{and}\ \bigl(\frac{1}{\hat{M}_{R}}\bigr)^{ij}
 \equiv \frac{1}{M_{R}^{i}}\delta^{ij}$. 

 Below the electroweak symmetry breaking scale, this dimension five operator generates
 the small left-handed Majorana neutrino mass matrix ${\cal M}_{\nu}$,
\begin{eqnarray}
({\cal M}_{\nu})^{\alpha \beta}\equiv m_{\nu}^{\alpha \beta}=\frac{v^{2}}{2}\Bigl[H^{\ast} \frac{1}{\hat{M}_{R}}
 H^{\dagger}\Bigr]^{\alpha \beta},  \label{mass}
 \end{eqnarray}
 which is a complex valued symmetric matrix. Here, $v$=246[GeV] is the VEV of the neutral higgs field, $\sqrt{2}\phi^{0}$.

\subsection{The condition to protect the primordial baryon asymmetry  \label{baryon}}
We require the condition that the primordial baryon asymmetry is not washed out.
 The sphaleron processes break the quantity, $(B+L)$, conserving the quantities, 
 $\bigl(\frac{B}{3}-L_{\alpha}\bigr)(\alpha=e, \mu, \tau)$, and are in equilibrium at the temperature between
 $100$[GeV] and $10^{12}$[GeV]. Here, we assume that all $M_{R}^{i}$ are within this temperature region.
 To protect the primordial baryon asymmetry, it is necessary that the lepton number violating processes are 
out of equilibrium through the temperature region,
\begin{eqnarray}
\Gamma_{\not{L}} < H,  \label{out}
 \end{eqnarray}
where $\Gamma_{\not{L}}$ is an interaction rate of a lepton number violating process and
 $H$ is the Hubble parameter, $H=1.66\sqrt{g_{\ast}}\frac{T^{2}}{M_{pl}} \simeq 1.44\times 10^{-18}\bigl(\frac{T^{2}}{[GeV]}\bigr)$ with $M_{pl}$ being the Plank mass.  Here, $g_{\ast}$ is the total degrees of freedom of effectively massless
particles.  In the seesaw model, we adopt the value $g_{\ast} = g_{\ast}^{SM} +g_{\ast}^{N_R}=$ 112
\footnote{This value is correct only when the most heaviest right-handed neutrino is in thermal equilibrium. Since 
we want to know the upper limits of the coupling constants, it is sufficient for us to adopt this value for all processes.}.
  In this model, the two kinds of lepton number violating processes exist. The first one is the decay and 
  inverse-decay of the heavy right-handed Majorana neutrinos, $N_R^{i} \leftrightarrow \nu_{L}^{\alpha} \phi^{0}$, and 
  the second kind is the scattering between two left-handed neutrinos and two neutral higgs,
  $\nu_{L}^{\alpha} \nu_{L}^{\beta} \leftrightarrow  \phi^{0 *}\phi^{0 *}$. 
  The out-of-equilibrium condition for each of these processes is examined as follows.
    \begin{itemize}
 \item $N_R$-decay and inverse-decay \par 
 We consider the processes, $N_R^{i} \leftrightarrow \nu_{L}^{\alpha} \phi^{0}.$ The interaction rates for these processes 
 are estimated at a temperature $T \gtrsim M_{R}^{i}$,
\begin{eqnarray}
\Gamma(N_{R}^{i} \leftrightarrow \nu_{L}^{\alpha} \phi^{0}) = \frac{1}{8\pi}|h^{\alpha i}|^{2}\frac{|M_{R}^{i}|^{2}}{T}.
 \end{eqnarray}
If we are going to impose the out-of-equilibrium conditions of the processes for all of indexes, i and $\alpha$,
the relations
\begin{eqnarray}
\Gamma(N_{R}^{i} \leftrightarrow \nu_{L}^{\alpha} \phi^{0}) < H \bigg|_{T=M_{R}^{i}}
\ \Leftrightarrow \ \  \frac{|h^{\alpha i}|^{2}}{|M_{R}^{i}|} < \frac{3.6\times 10^{-17}}{[\mbox{GeV}]}, \label{decay}
 \end{eqnarray}
 should be satisfied. The inequality (\ref{decay}) can be immediately rewritten as the upper bounds on the left-handed neutrino 
masses as
\begin{eqnarray}
|m_{\nu}^{\alpha \beta}| < \frac{v^{2}}{2}\sum_{i} \biggl|\frac{h^{\beta i}h^{\alpha i}}{M_{R}^{i}}  \biggr| 
=3.3\times 10^{-3}[\mbox{eV}]. \label{small}
 \end{eqnarray}
On the other hand, the recent neutrino oscillation experiments have shown 
that there exist mass-squared differences, relevant for solar neutrino oscillation $\Delta_{s}$
\cite{Valle-kam}, and the atmospheric neutrino oscillation $\Delta_{a}$ \cite{S-K2},
 
\begin{eqnarray}
 5.1 \times 10^{-5} [\mbox{eV}^2] &<& \Delta_{s} < 1.9 \times 10^{-4} [\mbox{eV}^2], \\
 1.6 \times 10^{-3} [\mbox{eV}^2] &<& \Delta_{a} < 4.0 \times 10^{-3} [\mbox{eV}^2].
 \end{eqnarray}
 
 The upper limit for all neutrino mass matrix elements (\ref{small}) can't satisfy these two mass squared differences and
 the condition we impose has no solution.
 Actually, even if we relax the condition and impose that at least one lepton flavor is conserved, 
the baryon number still can be protected, in proportion to the primordial value of the
 conserved lepton flavor. For example, we attempt to impose the electron-flavor ($L_{e}$) symmetry on the seesaw model. 
 In that case, $m_{\nu}^{e \alpha}(\alpha=e, \mu, \tau)$ should all vanish and the two of the three mixing angles to diagonalize 
 the neutrino mass matrix turn out to be zero. It is clear that this result  again contradicts with the recent 
 experimental data of the neutrino oscillation. 
 The situation is the same, even if we impose $L_{\mu}$ or $L_{\tau}$, instead.
 Thus, the last choice is to impose an approximate flavor
 symmetry on the model. An approximate symmetry means that the rates of all the processes breaking 
an exact symmetry satisfy the out-of-equilibrium condition (\ref{out}).
Thus, the presence of an approximate $L_{\alpha}$-flavor symmetry ($\alpha=e, \mu$ or $\tau$), leads to  
 \begin{eqnarray}
\Gamma_{i \alpha} < H \bigg|_{T=M_{R}^{i}} \ \mbox{for all of}\ i=1,2 \ \mbox{and} \ 3 \ \Leftrightarrow
 |m_{\nu}^{\alpha \alpha}| < 3.3\times 10^{-3}[\mbox{eV}]. \label{eout}
 \end{eqnarray}

 \item Scattering between two left-handed neutrinos and two neutral higgs \par
 We consider the scattering processes, $ \phi^{0 *}\phi^{0 *} \leftrightarrow \nu_{L}^{\alpha} \nu_{L}^{\beta}$
 induced by the dimension five operator in Eq. (\ref{lfive}).
 The interaction rates for these processes are estimated at the temperature $T \ll M_{R}^{i}$ and 
 the out of equilibrium conditions of these processes read as
   \begin{eqnarray}
\Gamma( \phi^{0 \ast} \phi^{0 \ast} \leftrightarrow \nu_{L}^{\alpha} \nu_{L}^{\beta})
 \simeq \frac{1}{\pi} \biggl|\frac{h^{\beta i}h^{\alpha i}}{M_{R}^{i}}  \biggr|^{2}T^{3} \simeq
 \frac{|m_{\nu}^{\alpha \beta}|^{2} }{v^{4}}T^{3}  < H,
 \end{eqnarray}
 for all choices of $\alpha, \beta$.
 This inequality can be rewritten as the one for the left-handed neutrino masses as 
  \begin{eqnarray}
|m_{\nu}^{\alpha \beta}| < 0.13  [\mbox{eV}].  \label{five}
 \end{eqnarray}
 It is found that these upper bounds for the elements of neutrino mass matrix are weaker than those obtained from
 the combination of the out-of-equilibrium conditions for $N_R$-(inverse-)decay and
 the data of WMAP and KAmLAND in the next section.
 We will verify the invalidity of this constraints in the end of the next section.
\end{itemize}

\subsection{Constraints from the results of WMAP  \label{W}}
We further constrain the mass matrixes, allowed from the arguments in \ref{baryon},
 by imposing a condition obtained from the recent results of WMAP.
The investigation into CMB by WMAP recently put the following upper bounds on 
the sum of the neutrino masses \cite{Wmap1,Wmap2,Han,Elg,Wmap3}, 
  \begin{eqnarray}
\Omega_{\nu}h^{2}=\frac{\sum_{i}|m_{i}|}{93.5[\mbox{eV}]} < 0.0076 \ 
\ \Leftrightarrow  \ \sum_{i}|m_{i}| < 0.71 [\mbox{eV}] , \label{map}
 \end{eqnarray}
 where $m_{i}$(i=1,2,3) is a small left-handed neutrino mass eigenvalue. 
 Hereafter, we assume that $H$ and $M_{R}^{i}$ are real matrixes. 
Since ${\cal M}_{\nu}$ is a real symmetric matrix, the inequality (\ref{map}) can be written as 
\footnote{This equation is correct even if
we transform  left-handed neutrinos by the two kinds of re-phasing, $P=\pm iI_{3 \times 3}$, 
defined as $\vec{\nu}_{L}^{m}\prime=P^{\ast}U^{\dagger}\vec{\nu}_{L}.$
The invariability of the trace of the neutrino mass matrix is discussed in \cite{Zee}.}
 \begin{eqnarray}
\biggl|\sum_{i}m_{i}\biggr|=|m_{\nu}^{ee}+m_{\nu}^{\mu \mu}+m_{\nu}^{\tau \tau}| < 0.71 [\mbox{eV}]. \label{sum}
 \end{eqnarray}
Here, we further assume that all $M_{R}^{i}$ are either positive or negative. Under this assumption, 
the inequality (\ref{sum}) can be written as 
 \begin{eqnarray}
|m_{\nu}^{ee}|, |m_{\nu}^{\mu \mu}|, |m_{\nu}^{\tau \tau}| < 0.71 [\mbox{eV}]  \label{notsum}
 \end{eqnarray}
Here, we define $\vec{p}_{\alpha}(\alpha=e, \mu, \tau)$ for the convenience of the following discussions as 
    \begin{eqnarray}
\vec{p}_{\alpha} \equiv \frac{v}{\sqrt{2}}\Biggl(\frac{h^{\alpha 1}}{\sqrt{M_{1}}}, \frac{h^{\alpha 2}}{\sqrt{M_{2}}}
, \frac{h^{\alpha 3}}{\sqrt{M_{3}}}\Biggr).
 \end{eqnarray}
 Using the $\vec{p}_{\alpha}$,  we can write the elements of the neutrino mass matrix in Eq. (\ref{mass}) 
 in a form of inner product of vectors, 
 \begin{eqnarray}
m_{\nu}^{\alpha \beta} = \vec{p}_{\alpha} \cdot\vec{p}_{\beta} \ .    \label{matrix}
 \end{eqnarray}
 The inequality (\ref{notsum}) can be written as 
   \begin{eqnarray}
 |\vec{p}_{\alpha}|^{2} < 0.71 [\mbox{eV}], \label{Wup}
 \end{eqnarray}
 where $\alpha=e, \ \mu$\ and $\tau$.
 We thus can determine the upper bounds of the all elements of the mass matrix 
 depending on the choice of the flavor symmetry
 from Eq.  (\ref{eout}),  (\ref{matrix}) and  (\ref{Wup}).
  \begin{itemize}
\item[(1)] Approximate $L_{e}$-symmetry \par
Eqs.  (\ref{eout}), (\ref{matrix}) and (\ref{Wup}) put the upper bounds on the elements of the mass matrix,
  \begin{eqnarray}
|m_{\nu}^{ee}| &=&|\vec{p}_{e}|^{2} < 3.3 \times 10^{-3} [\mbox{eV}] \equiv S, \\
|m_{\nu}^{e\mu}| &=&|\vec{p}_{e} \cdot \vec{p}_{\mu}| \leq |\vec{p}_{e}| |\vec{p}_{\mu}|
 < 4.8 \times 10^{-2} [\mbox{eV}] \equiv M,  \\
|m_{\nu}^{\mu \tau}| &=&|\vec{p}_{\mu} \cdot \vec{p}_{\tau}| \leq |\vec{p}_{\mu}| |\vec{p}_{\tau}|
 < 0.71 [\mbox{eV}] \equiv L.
 \end{eqnarray}
We similarly derive the upper bounds on the other elements and summarize these inequality in the form of mass matrix,
using S, M and L defined above,
   \begin{eqnarray}
  \left(\begin{array}{ccc}
                                    |m_{\nu}^{ee}| & |m_{\nu}^{\mu e}|  & |m_{\nu}^{\tau e}| \\
                                     |m_{\nu}^{e\mu}|   & |m_{\nu}^{\mu\mu}| & |m_{\nu}^{\mu\tau}| \\
									 |m_{\nu}^{e\tau}|  & |m_{\nu}^{\mu\tau}| & |m_{\nu}^{\tau\tau}|
									 \end{array}\right)
  < \left(\begin{array}{ccc}
                                    S & M  & M \\
                                     M   & L & L \\
									 M   & L & L
									 \end{array}\right). \label{emat}
 \end{eqnarray} 
 
  \item[(2)] Approximate $L_{\mu}$-symmetry  \par
 We can determine the upper bounds on all the elements in a similar way to that in the case (1), 
    \begin{eqnarray}
 {\cal M}_{\nu} < \left(\begin{array}{ccc}
                                    L & M  & L \\
                                     M   & S & M \\
									 L  & M & L
									 \end{array}\right). \label{mumat} 
 \end{eqnarray} 
 \item[(3)] Approximate $L_{\tau}$-symmetry   \par
 Similarly,
   \begin{eqnarray}
   {\cal M}_{\nu} < \left(\begin{array}{ccc}
                                    L & L  & M \\
                                     L   & L & M \\
									 M  & M & S
									 \end{array}\right). \label{taumat}
	\end{eqnarray}								 
  \end{itemize}

\section{Seesaw model confronts KamLAND data}
In the present section, we first  review the current status of the
results from neutrino oscillation experiments including the recently reported first KamLAND results.
We then require that the mass matrixes restricted in \ref{W} satisfy the current results of the neutrino oscillation experiments,
for each of the normal and the inverted  mass spectra including quasi-degenerate type.
 We next confirm that there really exists a parameter
set of the theory which leads to the allowed neutrino mass matrix in each case. We further check that
 the allowed mass matrixes satisfy the constraint from the neutrinoless double beta decay.  We finally summarize
 the results of the present section and calculate the resultant lepton and baryon numbers of the present universe.

\subsection{The present status of the results from the neutrino oscillation experiments}
We summarize the present status of neutrino oscillation experiments to make
clear what type of the neutrino mass matrix is allowed.  
The Super-Kamiokande Collaboration(S-K) shows that there exist a mass squared difference
$\Delta_{a}$ and a mixing angle $\theta_{atm}$ in order to explain the
atmospheric neutrino oscillation \cite{S-K,S-K2,Valle-Gon},
  \begin{eqnarray}
 1.6 \times 10^{-3} [\mbox{eV}^2] < \Delta_{a} < 4.0 \times 10^{-3} [\mbox{eV}^2], \ \ 
0.88 < \sin^2{2\theta_{atm}} \le 1.0  \ (90 \% \ \mbox{C.L.}),   \label{SK}
\end{eqnarray}
with the best fit values $\Delta_{a}=2.5 \times 10^{-3} [\mbox{eV}^2]$
and $\sin^2{2\theta_{atm}}=1.00$. 
The global analysis of the first results of KamLAND combined with the existing data of
solar neutrino experiments shows that there exist a mass squared
difference $\Delta_{s}$ and a mixing angle $\theta_{\odot}$ to account for the solar neutrino oscillation
\cite{SNO,kam,SNO2,kam2,Valle-kam},
\begin{eqnarray}
5.1 \times 10^{-5} [\mbox{eV}^2] < \Delta_{s} < 9.7 \times 10^{-5} [\mbox{eV}^2]&,& \  \ 
1.2 \times 10^{-4} [\mbox{eV}^2] < \Delta_{s} < 1.9 \times 10^{-4} [\mbox{eV}^2], \label{two} \nonumber
\\ 0.29 < \tan^2{\theta_{\odot}} <0.86 && (3 \sigma \ \mbox{level}), \label{non}
\end{eqnarray}
with the best fit values, $\Delta_{s}=6.9 \times 10^{-5}
[\mbox{eV}^2]$ and $\tan^2{\theta_{\odot}}=0.46$.
The CHOOZ experiment has put the upper bounds on the mixing angle $\theta_{13}$\cite{Chooz},
\begin{eqnarray}
\sin^2{2\theta_{13}} \le 0.1. \label{CH} 
\end{eqnarray}

Using the Maki-Nakagawa-Sakata(MNS) matrix U,
we can diagonalize a Majorana neutrino mass matrix ${\cal M}_{\nu}$ to $\hat{\cal M}_{\nu}$,
\begin{eqnarray}
 \hat{\cal M}_{\nu} =\left(\begin{array}{ccc}
                                    m_{1} & 0 & 0 \\
                                     0  & m_{2} & 0 \\
									 0 & 0 & m_{3}
									 \end{array}\right)=  U^{T}{\cal M}_{\nu}U,  \hspace{15mm}
 \vec{\nu}_{L}^{m} =  U^{\dagger} \vec{\nu}_{L}\equiv \left(\begin{array}{c}
    	\nu_{1}\\   
       \nu_{2}  \\ 
         \nu_{3}
	\end{array}\right)_{L},
\end{eqnarray}
where the weak eigenstates in $\vec{\nu}_{L}$ are understood to be the partners of the mass eigenstates of
charged leptons and $\vec{\nu}_{L}^{m}$ contain the neutrino mass eigenstates.
  We write a diagonal matrix with the symbol " $\hat{ }$ ", hereafter.
Since  we assume that ${\cal M}_{\nu}$ is a real symmetric matrix,
 we can parameterize the MNS matrix U as a orthogonal matrix,
\begin{eqnarray}
 U &\equiv&  
   \left(\begin{array}{ccc}
    1   &   0   & 0    \\   
    0   &   c_{23}   &  s_{23}     \\ 
   0   &   -s_{23}    &  c_{23}     
	\end{array}\right)
 \left(\begin{array}{ccc}
    c_{13}   &   0   & s_{13}    \\   
    0   &   1   &  0      \\ 
   -s_{13}   &   0    &  c_{13}     
	\end{array}\right)
 \left(\begin{array}{ccc}
    c_{12}   &   s_{12}   & 0    \\   
    -s_{12}   &   c_{12}  &  0      \\ 
   0  &   0    &  1     
	\end{array}\right)  \nonumber \\
	&=& R_{23}(-\theta_{23})R_{13}(\theta_{13})R_{12}(-\theta_{12}), 
\end{eqnarray}
where $c_{ij} \equiv \cos \theta_{ij}$ and $s_{ij} \equiv \sin \theta_{ij}$. 
Using these parameters, the two mass squared differences and two mixing angles in Eq. (\ref{SK}) and (\ref{non})
are written as 
 \begin{eqnarray}
\Delta_{a}=|m_{3}^{2}-m_{2}^{2}|, \hspace{5mm} \Delta_{s}=m_{2}^{2}-m_{1}^{2}, \hspace{10mm}
\theta_{atm}=\theta_{23}, \hspace{5mm} \theta_{\odot}=\theta_{12}.
 \end{eqnarray}
 The experimental data \cite{SNO,kam,SNO2,kam2,Valle-kam} for solar neutrino also show
 that the sign of $m_{2}^{2}-m_{1}^{2}$ is positive.
For brevity, we assume the mixing angle $\theta_{13}=0$, consistent with (\ref{CH}), 
and fix the two mass squared differences and ${\theta} _{atm}$ at the best fit values.
 Under these assumptions, we can write the matrix U using the only one 
 relatively poorly known parameter $\theta_{\odot}\equiv \theta$,
  \begin{eqnarray}
 U =\left(\begin{array}{ccc}
    \cos \theta   &   \sin \theta   & 0    \\   
    -\frac{1}{\sqrt{2}}\sin \theta    &   \frac{1}{\sqrt{2}}\cos \theta   &  \frac{1}{\sqrt{2}}     \\
   \frac{1}{\sqrt{2}}\sin \theta   &   -\frac{1}{\sqrt{2}}\cos \theta   &     \frac{1}{\sqrt{2}}
	\end{array}\right)
	 \hspace{5mm} \mbox{with}  \hspace{3mm}   0.29 < \tan^2\theta <0.86  \ (3 \sigma \ \mbox{level}).
 \end{eqnarray}
 
Then, the neutrino mass matrix ${\cal M}_{\nu}$ can be written as  
 \begin{eqnarray}
 {\cal M}_{\nu}=\left(\begin{array}{ccc}
    c^{2}m_{1}+s^{2}m_{2}   &  -\frac{1}{\sqrt{2}}sc(m_{1}-m_{2})    &   \frac{1}{\sqrt{2}}sc(m_{1}-m_{2})   \\   
      -\frac{1}{\sqrt{2}}sc(m_{1}-m_{2})  &   \frac{1}{2}(s^{2}m_{1}+c^{2}m_{2}+m_{3})  &  
	  -\frac{1}{2}(s^{2}m_{1}+c^{2}m_{2}-m_{3})   \\ 
   \frac{1}{\sqrt{2}}sc(m_{1}-m_{2})   &   -\frac{1}{2}(s^{2}m_{1}+c^{2}m_{2}-m_{3}) & 
   \frac{1}{2}(s^{2}m_{1}+c^{2}m_{2}+m_{3})
	\end{array}\right), \label{m}
 \end{eqnarray}
where  $c \equiv \cos \theta$ and $s \equiv \sin \theta$. Since we have fixed
 the values of the two mass squared differences, there remains only one independent parameter
 among three mass eigenvalue, $m_{1}, m_{2}, m_{3}$. In addition, the neutrino mass matrix
(\ref{m}) must satisfy the following two conditions: first, since  all $M_{R}^{i}$ are either
positive or negative, all $m^{ee}, m^{\mu\mu} \ \mbox{and} \  m^{\tau\tau}$
 are of the same sign. Second, the off diagonal elements must satisfy 
the following inequality,
   \begin{eqnarray}
|m_{\nu}^{e\mu}| &=&|m_{\nu}^{e\tau}| =|\vec{p}_{e} \cdot \vec{p}_{\mu}| \leq |\vec{p}_{e}| |\vec{p}_{\mu}|
=\sqrt{|m_{\nu}^{ee}||m_{\nu}^{\mu\mu}|},  \label{dot1}\\
|m_{\nu}^{\mu \tau}| &=&|\vec{p}_{\mu} \cdot \vec{p}_{\tau}| \leq |\vec{p}_{\mu}| |\vec{p}_{\tau}|
 < \sqrt{|m_{\nu}^{\mu\mu}||m_{\nu}^{\tau\tau}|}=|m_{\nu}^{\mu\mu}|.  \label{dot2}
 \end{eqnarray}

\subsection{Normal mass spectrum \label{Nor}}
We first investigate the normal mass spectrum, $|m_{1}| < |m_{2}| < |m_{3}|$.
Choosing $m_{1}$ as a parameter, we can express $m_{2}$ and $m_{3}$ as a function of $m_{1}$, 
 \begin{eqnarray}
m_{3}&=&\pm \sqrt{\Delta_{a}+\Delta_{s}+m_{1}^{2}}, \\
m_{2}&=&\pm \sqrt{\Delta_{s}+m_{1}^{2}}, \\
m_{1}&>&0,
 \end{eqnarray}
 where $m_{1}$ can be taken to be non-negative with no loss of generality 
 thanks to the freedom of the re-phasing, 
 $\vec{\nu}_{L}^{m}\prime=\pm i \vec{\nu}_{L}^{m}$.
 Every element of the neutrino mass matrix can be written as 
 \begin{eqnarray}
m^{ee}&=&c^{2}m_{1}\pm s^{2} \sqrt{\Delta_{s}+m_{1}^{2}}, \label{ee}  \\
m^{e\mu}&=&-m^{e\tau}=-\frac{1}{\sqrt{2}}sc\biggl(m_{1}\mp \sqrt{\Delta_{s}+m_{1}^{2}}\biggr) ,  \label{emu} \\
m^{\mu \tau}&=&-\frac{1}{2}\biggl(s^{2}m_{1}\pm c^{2}\sqrt{\Delta_{s}+m_{1}^{2}}\mp
\sqrt{\Delta_{a}+\Delta_{s}+m_{1}^{2}}\biggr), \label{mutau} \\
m^{\mu \mu}&=&m^{\tau \tau}=\frac{1}{2}\biggl(s^{2}m_{1}\pm c^{2}\sqrt{\Delta_{s}+m_{1}^{2}}\pm
\sqrt{\Delta_{a}+\Delta_{s}+m_{1}^{2}}\biggr). \label{mumu}
 \end{eqnarray}
 Since $m_{2}$ and $m_{3}$ can be either positive or negative, we must consider the four cases,
 (i) $m_{2}>0, m_{3}>0$, (ii)$m_{2}>0, m_{3}<0$, (iii)$m_{2}<0, m_{3}>0$ and (iv)$m_{2}<0, m_{3}<0$,
 to see if some approximate symmetries remain. It is easily found that the approximate $L_{\mu}$ or 
 $L_{\tau}$-symmetry is not possible.
 In the case (i), (iii) and (iv), since $|m_{\mu\mu}|$ is a monotonically increasing function of $m_{1}$ and goes to 
 infinity as $m_{1}$ goes to infinity, $|m_{\mu\mu}|$ is minimum at $m_{1}$=0. The following relation holds
  \begin{eqnarray}
|m_{(\mbox{i})}^{\mu\mu}|=|m_{(\mbox{iv})}^{\mu\mu}| \ge |m_{(\mbox{iii})}^{\mu\mu}|=
\frac{1}{2}\biggl(-c^{2}\sqrt{\Delta_{s}}+\sqrt{\Delta_{a}+\Delta_{s}}\biggr) \ge 2.2 \times 10^{-2} \mbox{[eV]},
  \end{eqnarray}
  at $m_{1}$=0. This inequality is inconsistent with (\ref{eout}). In the case (ii), 
  the condition (\ref{eout}) is applied to the Eq. (\ref{mumu}),
 \begin{eqnarray}
m_{\nu}^{\mu\mu}=m_{\nu}^{\tau\tau} \le 3.3 \times 10^{-3} \mbox{[eV]} 
 \Rightarrow m_{1} \ge 1.9 \times 10^{-1} \mbox{[eV]}  \Rightarrow |m_{\nu}^{\mu\tau}| \ge 1.9 \times 10^{-1}
 \mbox{[eV]}.  
 \end{eqnarray}
 This inequality is inconsistent with $|m_{\nu}^{\mu\tau}| \le 3.3 \times 10^{-3} \mbox{[eV]}$ in (\ref{dot2}).
 Thus, the approximate $L_{e}$-symmetry is the only remaining possibility, which we examine for the four cases.
  \begin{itemize}
\item[(i)] $m_{1} > 0, m_{2} > 0, m_{3} > 0$ \par
The condition (\ref{eout}) applied for the Eq. (\ref{ee})  yields for $0.29 < \tan^2{\theta} <0.86$ and 
$\Delta_{s}=6.9 \times 10^{-5} [\mbox{eV}^2]$,
 \begin{eqnarray}
0 < m_{1} \le 1.8 \times 10^{-3} \mbox{[eV]}
\Rightarrow 1.9 \times 10^{-3} \le m_{\nu}^{ee} \le 3.3 \times 10^{-3} \mbox{[eV]}.  \label{(i)}
 \end{eqnarray}
 Applying this range of $m_{1}$ to Eq. (\ref{emu}), (\ref{mutau}) and (\ref{mumu}),
 we can determine the remaining elements of the mass matrix 
 rather precisely with $\Delta_{a}=2.5 \times 10^{-3} [\mbox{eV}^2]$,
  \begin{eqnarray}
  &(&m_{\nu}^{e\mu}=-m_{\nu}^{e\tau}, m_{\nu}^{\mu \tau},  m_{\nu}^{\mu \mu}=m_{\nu}^{\tau\tau})  \nonumber \\
  &=&\Bigl((2.0\sim 2.9) \times 10^{-3}, (2.2 \sim 2.3) \times 10^{-2}, 
  (2.8 \sim 2.9) \times 10^{-2}   \Bigr)\mbox{[eV]}.\label{m1}
   \end{eqnarray}

\item[(ii)]$m_{1} > 0, m_{2} > 0, m_{3}<0$  \par
The allowed range, necessary for the approximate $L_{e}$-symmetry, is same as Eq. (\ref{(i)}) 
and yields 
 \begin{eqnarray}
 m_{\nu}^{\mu\mu}=m_{\nu}^{\tau\tau}<0,
 \end{eqnarray}
while $m_{\nu}^{ee}>0$, which contradicts with the condition that all $m_{\nu}^{ee},
 m_{\nu}^{\mu\mu}\ \mbox{and} \  m_{\nu}^{\tau\tau}$
 must be of the same sign. Hence, this case is excluded.

\item[ (iii)]$m_{1} > 0, m_{2} < 0, m_{3}>0$ \par
Now, the condition (\ref{eout}) reads as,
 \begin{eqnarray}
0<m_{1} \le 4.6 \times 10^{-2} \mbox{[eV]} \ \Rightarrow \ m_{\nu}^{ee} \le 3.3 \times 10^{-3} \mbox{[eV]}. \label{(iiii)}
 \end{eqnarray}
 Applying this range of $m_{1}$ to Eq. (\ref{emu}), (\ref{mutau}), (\ref{mumu}),  
  and imposing (\ref{dot1}) and (\ref{dot2}), we can determine the matrix elements as,
  \begin{eqnarray}
  &(&m_{\nu}^{e\mu}=-m_{\nu}^{e\tau}, m_{\nu}^{\mu \tau},  m_{\nu}^{\mu \mu}=m_{\nu}^{\tau\tau})  \nonumber \\
  &=&\Bigl( -1.0 \times 10^{-2} \sim -2.4 \times 10^{-3} , (2.7 \sim 3.2) \times 10^{-2}, 
     (2.1 \sim 3.2) \times 10^{-2} \Bigr)\mbox{[eV]}.\label{m2}
   \end{eqnarray}

\item[ (iv)]$m_{1} > 0, m_{2} < 0, m_{3}<0$ \par
The condition  (\ref{eout}) leads to the same condition as Eq. (\ref{(iiii)})
 \begin{eqnarray}
0<m_{1} \le 4.6 \times 10^{-2}  \mbox{[eV]} \ \Rightarrow \ m_{\nu}^{ee} \le 3.3 \times 10^{-3}  \mbox{[eV]} .
 \end{eqnarray}
 Following the same steps as in the case (iii), we get
  \begin{eqnarray}
  &(&m_{\nu}^{e\mu}=-m_{\nu}^{e\tau},  m_{\nu}^{\mu \tau},  m_{\nu}^{\mu \mu}=m_{\nu}^{\tau\tau})   \\
  &=& \Bigl( -1.2 \times 10^{-2} \sim -2.4 \times 10^{-3}, -(2.1 \sim 3.2) \times 10^{-2}, 
  -(2.7 \sim 4.7) \times 10^{-2}  \Bigr)\mbox{[eV]}.\nonumber \label{m3}
   \end{eqnarray} 
\end{itemize}
We should show that there really exists a parameter set of the seesaw model, $h^{\alpha i}$ and $M_{R}^{i}$,
 which induces such determined neutrino mass matrix
for each of the three cases. But, since the proof is rather a routine work, we execute it in the APPENDIX \ref{ex}.

\subsection{Inverted mass spectrum}
We next investigate the inverted mass spectrum, $|m_{3}| < |m_{1}| < |m_{2}|$.
Choosing the smallest mass $m_{3}$ as a parameter, we can express $m_{1}$ and $m_{2}$ as a function of $m_{3}$, 
 \begin{eqnarray}
m_{2}&=&\pm \sqrt{\Delta_{a}+\Delta_{s}+m_{3}}, \\
m_{1}&=&\pm \sqrt{\Delta_{a}+m_{3}}, \\
m_{3} &>&0,
 \end{eqnarray}
 where $m_{3}$ can be taken to be non-negative without loss of generality.
  Every element of the neutrino mass matrix can be written as 
  \begin{eqnarray}
m^{ee}&=&\pm c^{2}\sqrt{\Delta_{a}+m_{3}^{2}}\pm s^{2}\sqrt{\Delta_{a}+\Delta_{s}+m_{3}^{2}}, \label{iee} \\
m^{e\mu}&=&-m_{e\tau}=-\frac{1}{\sqrt{2}}sc\biggl(\pm \sqrt{\Delta_{a}+m_{3}^{2}}
\mp  \sqrt{\Delta_{a}+\Delta_{s}+m_{3}^{2}} \biggr), \label{iemu}\\
m^{\mu \tau}&=&-\frac{1}{2}\biggl(\pm s^{2}\sqrt{\Delta_{a}+m_{3}^{2}}\pm
 c^{2}\sqrt{\Delta_{a}+\Delta_{s}+m_{3}^{2}}-m_{3}\biggr), \label{imutau}\\
m^{\mu \mu}&=&m^{\tau \tau}=\frac{1}{2}\biggl(\pm s^{2}\sqrt{\Delta_{a}+m_{3}^{2}}
 \pm c^{2}\sqrt{\Delta_{a}+\Delta_{s}+m_{3}^{2}}+m_{3}\biggr). \label{imumu}
 \end{eqnarray} 
 We must consider the four cases, (i) $m_{1}>0, m_{2}>0$, (ii)$m_{1}>0, m_{2}<0$, 
 (iii)$m_{1}<0, m_{2}>0$ and (iv)$m_{1}<0, m_{2}<0$. 
  It is easily found that the approximate $L_{e}$-symmetry is not allowed. 
 Since $|m^{ee}|$ is a monotonically increasing function of $m_{3}$ and goes to 
 infinity as $m_{3}$ goes to infinity, $|m^{ee}|$ is minimum at $m_{3}$=0. The following relation is satisfied
  \begin{eqnarray}
|m_{(\mbox{i})}^{ee}|=|m_{(\mbox{iv})}^{ee}| \ge |m_{(\mbox{ii})}^{ee}|=|m_{(\mbox{iii})}^{ee}|
=c^{2}\sqrt{\Delta_{a}}-s^{2}\sqrt{\Delta_{a}+\Delta_{s}} \ge 3.6\times 10^{-3} \mbox{[eV]},
  \end{eqnarray}
at $m_{3}$=0. This inequality is inconsistent with (\ref{eout}). 
 Thus, we examine only the approximate $L_{\mu}$ or $L_{\tau}$-symmetry,
 as remaining possibilities, though we find these symmetries are not realized.
  \begin{itemize}
\item[(i)] $m_{3} > 0, m_{1} > 0, m_{2} > 0$ \par
Since the elements $m_{\nu}^{\mu\mu}$ or $m_{\nu}^{\tau\tau}$ is too large 
to satisfy Eq. (\ref{eout}) for an arbitrary $m_{3}$, this mass matrix
does not possess an approximate $L_{\mu}$ nor $L_{\tau}$-symmetry.

\item[(ii)] $m_{3} > 0, m_{1} > 0, m_{2} < 0$ \par
The condition (\ref{eout}) is applied to the Eq. (\ref{imumu}),
 \begin{eqnarray}
 4.3\times 10^{-3} \mbox{[eV]} \le m_{3} \le 4.4 \times 10^{-2} \mbox{[eV]}, 
 \end{eqnarray}
which, in tern, leads through Eq. (\ref{imutau}) to 
 \begin{eqnarray}
 m_{\nu}^{\mu \tau} \ge 4.3 \times 10^{-3} \mbox{[eV]}.  \nonumber
 \end{eqnarray}
 This value is inconsistent with (\ref{dot2}): $|m_{\nu}^{\mu\tau}| \le 3.3 \times 10^{-3} \mbox{[eV]}$.

 \item[(iii)] $m_{3} > 0, m_{1} < 0, m_{2} > 0$  \par
Eq. (\ref{iee}) and (\ref{imumu}) determine the sign of $m_{\nu}^{ee}$ and $m_{\nu}^{\mu\mu}$, 
 \begin{eqnarray}
m_{\nu}^{ee}<0, \ m_{\nu}^{\mu\mu}=m_{\nu}^{\tau\tau}>0.
 \end{eqnarray}
  This is inconsistent with the condition that all $m_{\nu}^{ee}, m_{\nu}^{\mu\mu}\ \mbox{and} \  m_{\nu}^{\tau\tau}$
 must be of the same sign.

 \item[(iv)] $m_{3} > 0, m_{1} < 0, m_{2} < 0$ \par

The condition (\ref{eout}) is applied to the Eq. (\ref{imumu}),
 \begin{eqnarray}
m_{\nu}^{\mu\mu}=m_{\nu}^{\tau\tau} \le 3.3 \times 10^{-3} \mbox{[eV]} 
 \Rightarrow m_{3} \ge 1.9 \times 10^{-1} \mbox{[eV]}  \Rightarrow |m_{\mu\tau}| \ge 1.9 \times 10^{-1}\mbox{[eV]}. 
 \end{eqnarray}
 This inequality is inconsistent with $|m_{\nu}^{\mu\tau}| \le 3.3 \times 10^{-3} \mbox{[eV]}$ in (\ref{dot2}).

 \end{itemize}

%\subsection{Existence proof of the parameters}
%Here, we show that there exists a parameter set of the seesaw model which induces an allowed neutrino mass matrix.
% We find a set of parameter for each mass matrix allowed in \ref{Nor}.

\subsection{Experimental constraint on $0\nu\beta\beta$ amplitude}
The amplitude of the neutrinoless double beta decay including the exchange of Majorana neutrinos 
with Majorana mass insertion due to $W^{\pm}$-exchange is proportional to $m_{ee}$
 as was first pointed out by Wolfenstein \cite{Wol2}. The absence of this process so far 
 reported gives the upper bounds on the element $m_{ee}$ in the neutrino mass matrix
\cite{Particle}\footnote{It is recently reported that the first
evidence for 0$\nu\beta \beta$ is observed\cite{0}. The value is
$|m_{ee}|$=0.11 $\sim$ 0.56 eV (95 $\%$ C.L.) with the best fit value
0.39 eV.  There are, however, arguments against this report, too
\cite{Aal}.},
\begin{eqnarray}
  |m_{\nu}^{ee}| < 0.1 \ \mbox{[eV]}  \ (90 \% \ \mbox{C.L.}). \label{mee}
\end{eqnarray}
The values obtained in the allowed neutrino mass matrixes, coming from (\ref{eout}),
\begin{eqnarray}
  |m_{\nu}^{ee}| \le 3.3 \times 10^{-3} \ \mbox{[eV]},
\end{eqnarray}
are all consistent with the bounds in Eq. (\ref{mee}).

\subsection{Allowed neutrino mass matrix and final baryon number}
We summarize what types of the neutrino mass matrixes are allowed in the present section on the TABLE \ref{allow}.  
\begin{table}[ht] \begin{tabular}{|c|c|c|c|}\hline
Flavor symmetry & (1)$L_{e}$  & (2)$L_{\mu}$ & (3)$L_{\tau}$ \\ \hline 
Normal spectrum  & \multicolumn{3}{c|}{}   \\ \hline 
(i) $m_{1} > 0, m_{2} > 0, m_{3} > 0$ & $\bigcirc$ & $\times$ &  $\times$   \\ \hline 
(ii) $m_{1} > 0, m_{2} > 0, m_{3} < 0$ & $\times$  & $\times$  & $\times$  \\ \hline 
(iii) $m_{1} > 0, m_{2} < 0, m_{3} > 0$ & $\bigcirc$ & $\times$  & $\times$  \\ \hline 
(iv) $m_{1} > 0, m_{2} < 0, m_{3} < 0$ & $\bigcirc$ & $\times$  & $\times$  \\ \hline 
 Inverted spectrum  & \multicolumn{3}{c|}{}   \\ \hline 
(i) $m_{3} > 0, m_{1} > 0, m_{2} > 0$ &  $\times$ &  $\times$ &  $\times$ \\ \hline 
(ii) $m_{3} > 0, m_{1} > 0, m_{2} < 0$ &  $\times$ &  $\times$ &  $\times$ \\ \hline 
(iii) $m_{3} > 0, m_{1} < 0, m_{2} > 0$ &  $\times$ &  $\times$ & $\times$  \\ \hline 
(iv) $m_{3} > 0, m_{1} < 0, m_{2} < 0$ &  $\times$ &  $\times$ &  $\times$ \\ \hline 
\end{tabular}
\caption{The allowed neutrino mass matrixes are shown.
The approximate $L_{e}$-symmetry in the the normal mass spectrum is only allowed.  \label{allow}}
\end{table}
The $L_{e}$-symmetry in the normal spectrum is only allowed.
Since the square of the lightest mass eigenvalue, $m_{1}$, is very small compared with the solar mass 
squared deference in the case (i) and are equal to the atmospheric mass squared deference in the case (iii),(iv),
the neutrino mass spectrum is normal hierarchical in the case (i) and is slightly normal hierarchical in the case (iii),(iv), 
respectively.
It is also found that the inverted mass spectrum is not allowed.
We summarize the allowed neutrino mass matrixes in each of the three cases.
  \begin{itemize}
\item[(i)]$m_{1} > 0, m_{2} > 0, m_{3} > 0$
{\footnotesize
  \begin{eqnarray}
 {\cal M_{\nu}}=\left(\begin{array}{ccc}
    (1.9\sim 3.3) \times 10^{-3}   &   \cdots   &  \cdots \\ 
     (2.0\sim 2.9) \times 10^{-3}   &  (2.8 \sim 2.9) \times 10^{-2}      &  \cdots    \\ 
   -(2.0 \sim 2.9) \times 10^{-3}  &   (2.2 \sim 2.3) \times 10^{-2}     &      (2.8 \sim 2.9) \times 10^{-2}
	\end{array}\right)\mbox{[eV]}. \label{al1}
 \end{eqnarray}}

\item[(iii)]$m_{1} > 0, m_{2} < 0, m_{3} > 0$
{\footnotesize
  \begin{eqnarray}
{\cal M_{\nu}}=\left(\begin{array}{ccc}
    (0 \sim 3.3) \times 10^{-3}   &  \cdots   &  \cdots \\   
     -1.0\times 10^{-2} \sim  -2.4\times 10^{-3}   &  (2.1\sim 3.2) \times 10^{-2}      &    \cdots  \\ 
    2.4\times 10^{-3} \sim 1.0\times 10^{-2}  &   (2.7 \sim 3.2) \times 10^{-2}     &      (2.1 \sim 3.2) \times 10^{-2}
	\end{array}\right)\mbox{[eV]}.  \label{al2}
 \end{eqnarray}}
\item[(iv)]$m_{1} > 0, m_{2} < 0, m_{3} < 0$
{\footnotesize
 \begin{eqnarray}
 {\cal M_{\nu}}=\left(\begin{array}{ccc}
    (-3.3 \sim 0) \times 10^{-3}   &   \cdots  &   \cdots \\   
     -1.2\times 10^{-2}  \sim   -2.4\times 10^{-3}   &  -(2.7 \sim 4.7) \times 10^{-2}      &   \cdots   \\ 
    2.4\times 10^{-3} \sim 1.2\times 10^{-2}  &   -(2.1 \sim 3.2) \times 10^{-2}     &      -(2.7 \sim 4.7) \times 10^{-2}
	\end{array}\right)\mbox{[eV]}.  \label{al3}
 \end{eqnarray}}
 \end{itemize}
 These mass matrixes satisfy the constraints from the processes induced through the dimension five operator (\ref{five}).
When the sphaleron processes are in equilibrium, the quantity $(\frac{B}{3}-L_{e})$ is approximately conserved.
Hence, the primordial value, $(\frac{B}{3}-L_{e})_{initial}$, is divided into the baryon and lepton numbers 
in the present universe denoted by $B_{final}$ and $L_{final}$ at the following rate \cite{Hav},
 \begin{eqnarray}
B_{final}=-\frac{28}{51}L_{final}=\frac{84}{277}\biggl(\frac{B}{3}-L_{e} \biggr)_{final}
=\frac{84}{277}\biggl(\frac{B}{3}-L_{e} \biggr)_{initial}. 
 \label{pro}
 \end{eqnarray}

\section{Summary}
We discussed the condition that the lepton number violating processes in
the seesaw model do not wash out the primordial baryon or lepton asymmetry. In 
addition to this condition, we also required the conditions on the neutrino masses and mixings imposed by
the recent results of WMAP and the neutrino oscillation experiments including the first results of KamLAND. 
We investigated whether these exist some solutions in the seesaw
 model which satisfy all of these conditions or not. As a result, it was found that the neutrino mass matrixes with
 the approximate $L_{e}$-symmetry in the normal hierarchy of the neutrino masses are the unique possibilities.
 This result indicates that the neutrino oscillation experiments and WMAP favor the approximate 
 $L_{e}$-symmetry in the normal hierarchy.
  We could determine the allowed neutrino mass matrixes with accuracy in (\ref{al1}), (\ref{al2}) and (\ref{al3}).
 We confirmed that there exists a parameter set of the seesaw model which leads to the allowed mass matrix
 for each case. 
  We also checked that these allowed mass matrixes all satisfy the constraint from the neutrinoless double beta decay. 
We finally calculated the present baryon and lepton numbers in terms of the primordial value of $(\frac{B}{3}-L_{e})$.

\begin{acknowledgments}
I would like to thank C. S. Lim for helpful and fruitful discussions and for modifying the present paper.  
I also thank K. Ogure for valuable comments on this manuscript.
\end{acknowledgments}

\appendix
\section{Existence proof of the parameters \label{ex}}
Here, we show that there exists a parameter set of the seesaw model which induces an allowed neutrino mass matrix.
 We find a set of parameter for each allowed mass matrix with the approximate $L_{e}$-symmetry discussed
 in \ref{Nor}.
 \begin{itemize}
\item[(i)] $m_{1} > 0, m_{2} > 0, m_{3} > 0$  \par
As our purpose is the existence proof, to simplify the analysis we demand 
\begin{eqnarray}
 h^{e1}=h^{\mu 1}=h^{\tau 1}=0, \  p_{e2}=p_{e3}, \ p_{\mu 1}=p_{\tau 1} \ \mbox{and} \ p_{\mu 3}=-p_{\tau 2}.
 \end{eqnarray}
 We find that the values chosen from  (\ref{al1}) 
 \begin{eqnarray}
(m^{ee}, m^{e\mu}, m^{\mu \tau}, m^{\mu \mu})=(2.0 \times 10^{-3}, 2.3 \times 10^{-3},
 2.3 \times 10^{-2}, 2.8 \times 10^{-2})\mbox{[eV]}.
 \end{eqnarray}
 are provided by a set of parameters,
 \begin{eqnarray}
\Bigl((p_{e2})^{2}, (p_{\mu 3})^{2}, (p_{\mu 1})^{2} \Bigr)=(1.0 \times 10^{-3}, 5.3 \times 10^{-3},
 2.3 \times 10^{-2} )\mbox{[eV]}.
 \end{eqnarray}
It is proved that there exists a parameter set which can induce an allowed neutrino mass matrix (\ref{al1}).
\item[(iii)] $m_{1} > 0, m_{2} < 0, m_{3} > 0$  \par
Requiring $h^{e1}=0,\  p_{e2}=p_{e3},\  p_{\mu 1}=p_{\tau 1}$ and $p_{\mu 2}=-p_{\tau 2}$,
 we find 
 \begin{eqnarray}
(m^{ee}, m^{e\mu}, m^{\mu \tau}, m^{\mu \mu})= (2.2 \times 10^{-3},
 -2.4 \times 10^{-3}, 3.2 \times 10^{-2}, 3.0 \times 10^{-2})\mbox{[eV]} \nonumber 
 \end{eqnarray}
is satisfied by 
 \begin{eqnarray}
\Bigl((p_{e2})^{2}, (p_{\mu 1})^{2}, (p_{\mu 2})^{2} \Bigr)=(1.1 \times 10^{-3},
 3.0 \times 10^{-2}, 1.4 \times 10^{-3} )\mbox{[eV]}. 
 \end{eqnarray}

\item[(iv)] $m_{1} > 0, m_{2} < 0, m_{3} < 0$  \par
For $h^{e2}=h^{\mu 3}=h^{\tau 3}=0, \ p_{\mu 1}=-p_{\tau 1}$ and $p_{\mu 2}=p_{\tau 2}$,
 \begin{eqnarray}
(m^{ee}, m^{e\mu}, m^{\mu \tau}, m^{\mu \mu})=
(-1.0 \times 10^{-3}, -2.5 \times 10^{-3},   -2.2 \times 10^{-2}, -4.5 \times 10^{-2})\mbox{[eV]} \nonumber
 \end{eqnarray}
   is given by 
   {\small
 \begin{eqnarray}
\Bigl((p_{e1})^{2}, (p_{e 3})^{2}, (p_{\mu 1})^{2}, (p_{\mu 2})^{2} \Bigr)= 
(5.4 \times 10^{-4}, 4.6 \times 10^{-4}, 1.2 \times 10^{-2}, 3.4 \times 10^{-2} ) \mbox{[eV]}.
 \end{eqnarray}}
 \end{itemize}

\end{document}